\documentclass[twocolumn,prb,showpacs,floatfix,superscriptaddress]{revtex4}  
\usepackage{graphicx}
\usepackage{amssymb}

\def\beeq{\begin{equation}}
\def\eneq{\end{equation}}
\def\beeqa{\begin{eqnarray}}
\def\eneqa{\end{eqnarray}}

\begin{document}
\DeclareGraphicsExtensions{.pdf,.ps,.eps}

 \title{
 Sampling the diffusion paths of a neutral vacancy in Silicon with
 SIEST-A-RT}
 
\author{Fedwa El-Mellouhi}
\email {f.el.mellouhi@umontreal.ca}
\author{Normand Mousseau }
\email {Normand.Mousseau@umontreal.ca}
\affiliation{D\'epartement de physique and Regroupement  qu\'eb\'ecois sur les mat\'eriaux de pointe, Universit\'e de  Montr\'eal, C.P. 6128, succ. Centre-ville, Montr\'eal (Qu\'ebec) H3C  3J7, Canada}

\author{Pablo  Ordej\'on}
\affiliation{Institut de Ci\`encia de Materiales de Barcelona (CSIC),  Campus de la Universitat Aut\'onoma de Barcelona, Bellaterra,   E-08193, Barcelona, Spain.}

\date{\today}

\begin{abstract}
We report a first-principles study of vacancy-induced self-diffusion
in crystalline silicon. Starting form a fully relaxed configuration
with a neutral vacancy, we proceed to search for local diffusion
paths. The diffusion of the vacancy proceeds by hops to first nearest
neighbor with an energy barrier of 0.40~eV in agreement with 
experimental results. Competing
mechanisms are identified, like the reorientation, and the
recombination of dangling bonds by Wooten-Winer-Weaire process.
\end{abstract}


\maketitle
\section{Introduction}
\label{sec:intro}

Self-diffusion in homogeneous solids is a fundamental process of mass
transport, in addition, it is responsible of the annealing of 
implantation damage, crystal to amorphous transition and the
nucleation of extended defects. The native defects, such as vacancies
and self-interstitial, have been identified as the prime entities
controlling the self-diffusion, as well as the diffusion of impurities
in semiconductor lattices. For example, in silicon, antimony
impurities are vacancy diffusers and phosphorus are interstitial
diffusers~\cite{URAL99}. Therefore, self-diffusion in Si
has been extensively investigated by experiments~\cite{ Ma99, WATK92,
  Corbett64,Danne86,URAL99} and simulations~\cite{ Ghaisas91, Cheli99,
  Colom01, Vech89, Lewis96, Puska98, AN98, Yam84, Kum01, Ala93, Goe02,
  AM01, Kim00, WA91, Tang97, Le99, ES00, Hen02, Mu99, Pu00,Me98,
  PROB03, UBER00,ROBER94, KIM97, ANDER96,Clarck97, Len03}. 

In covalently bonded silicon, the neutral vacancy is  the most stable charge state 
near room temperature ~\cite{WATK92, Lopez88}.  While the
diffusion mechanism and the barrier for silicon have been
experimentally determined decades ago~\cite{Danne86,Corbett64}, we
still lack a detailed knowledge of the various activated mechanisms
associated with the neutral vacancy.  With standard {\it ab initio} molecular dynamics methods, the number of events generated is always insufficient to give
reliable data for migration energies. Accordingly, most of the focus
has been on formation energies, distortion and reorientation
geometries~\cite{ANDER96}. Few papers reported a calculated activation
energy and a migration path that are in agreement with the
experimental data~\cite{ROBER94, Maroud93}.  Until now, the only
studies available for exploring the energy landscape were that of
Munro and Wales~\cite{Mu99} who examined the dynamics of a vacancy,
among others, using a tight-binding approach, and the work of Kumeda
{\it et al.} ~\cite{Kum01} using first-principles calculations. They identified the
migration barrier and the underlying mechanism. However, a complete
study for the topology of the landscape around a stable minimum is still needed.

The principal goal of this study is to give a
complete description of the diffusion mechanism in elementary
semiconductors. As such, the SIEST-A-RT method is built up to develop an entire
methodology for quantum mechanically precise and reasonably fast
energy landscape exploration method.  Our simulations are performed on
supercells containing 215 atoms. This supercell size is
computationally tractable and constitutes a reasonable size to
simulate the host crystal. We generate the diffusion paths using the
activation-relaxation technique (ART)~\cite{MOU98, BAR96}, which can
sample efficiently the energy landscape of complex systems. The forces
and energies are evaluated using SIESTA~\cite{SAN97,Soler02}, a
self-consistent density functional method using standard
norm-conserving pseudo-potentials and a flexible numerical linear
combination of atomic orbital basis set. Combining these two methods
allows us to identify very efficiently diffusion paths of various energy height.

This paper is constructed as follows. In the next section, we discuss
the SIEST-A-RT method; simulation details are given in section
\ref{sec:detail}. We apply the method to the vacancy in silicon, the
corresponding activated dynamics are detailed and discussed in section
\ref{sec:Res}. Finally we emphasize the capability of the method in
identifying competing events in silicon and its possible extention  for
future applications.

\section{Method: SIEST-A-RT}

In order to sample the diffusion paths associated with a vacancy in
bulk Si, we integrate the activation-relaxation technique (ART
nouveau)~\cite{BAR96,Malek_ART} to the {\it ab-initio} program
SIESTA. This approach is similar to the integration of the hybrid
eigenvector-following method to a tight-binding and an {\it ab-initio} code
as performed by Munro, Kumeda and Wales~\cite{Mu99,Kum01} as well as the dimer-method~\cite{Hen99} which was coupled with VASP\cite{vasp} by Henkelman and J\`onsson\cite{Hen02}.

As both ART nouveau and SIESTA are described elsewhere in the
literature~\cite{Malek_ART,Soler02}, we review these methods only briefly. 

\subsection{ART nouveau}
\label{sec:ART}

The activation-relaxation technique is a method for sampling
efficiently the energy landscape of activated
systems.~\cite{BAR96,MOU98} Instead of following in details the
thermal fluctuations, it searches directly for transition states;
these are characterized as first-order saddle points in the energy
landscape. This method was refined recently to ensure a controlled
convergence to the saddle point (ART nouveau).~\cite{Malek_ART} An ART
event can be decomposed into the following steps.  {\it Activation:}
(1) Starting from a local minimum, push the configuration in a
direction chosen randomly in the 3N-dimensional space; stop when the
lowest eigenvalue becomes negative or when the configuration is far
enough from the initial configuration. (2) Follow the eigendirection corresponding to this negative eigenvalue while minimizing the energy in the perpendicular directions; stop when the total force (parallel and perpendicular to the eigendirection) falls below a given threshold ($0.1$ eV/\AA). This is the transition state. {\it Relaxation:}
After pushing over the saddle point, minimize the energy using any
standard minimization algorithm.

In order to sample only the activated mechanisms associated with the
vacancy, we restrict the initial random projection for each event to
the atoms surrounding the defect.  Except for this initial selection, all atoms are allowed
to move during the ART event. 

ART uses the Lancz\`os~\cite{Lan88} algorithm to extract the lowest eigenvalue and its corresponding eigenvector; 16 to 20 force evaluations are sufficient, irrespective of the size of the system, to extract a stable eigenvalue value and eigenvector.  Recent unpublished tests indicate that the number of force evaluations needed to reach a saddle point is about the same for the dimer method and ART nouveau~\cite{Hen99}, indicating no significant difference between the various activation-relaxation methods. 

\subsection{Siesta}

Forces and energies are evaluated using SIESTA~\cite{SAN97, Soler02}, a
self-consistent density functional method using standard
norm-conserving pseudo-potentials of Troullier-Martins~\cite{Trou91}
factorized in the Kleiman-Baylander form~\cite{Klei82}. Matrix
elements are evaluated on a 3D grid in real space.  The one-particle
problem is solved using linear combination of pseudo-atomic orbitals
(PAO) basis set of finite range. The quality of the basis can be
improved by {\it radial} flexibilization, add more than one radial
function within the same angular momentum (Multiple-$\zeta$), and {\it
  angular} flexibilization or polarization, add shells of different
atomic symmetry (different l).

Because ART requires a numerical derivative of the force to compute
the curvature of the energy landscape, it is essential to obtain
highly converged forces. The various parameters used are discussed
below.

\subsection{Optimization of the parameters}
\begin{figure}
\centerline{\includegraphics[width=7.5cm, angle=-90]{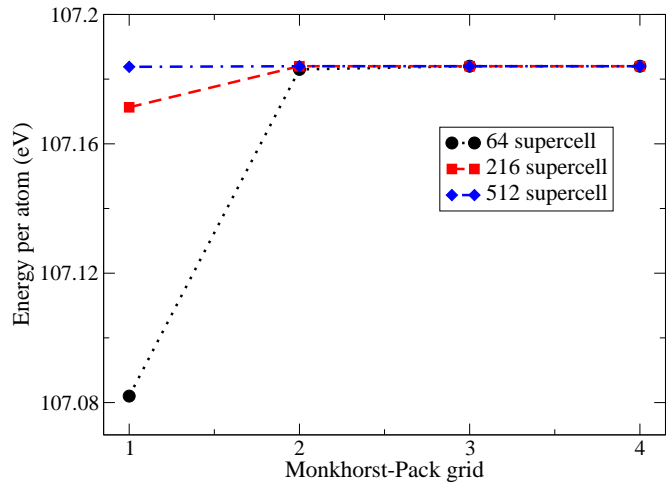}}
\caption{"(Color on line)" Convergence tests for the energy per atoms as a function of  Brillouin zone sampling for 64, 216 and 512 cubic supercells. The lines are guide to the eye. (see the text).  }
\label{fig:bz}
\end{figure}

The introduction of a vacancy in the crystal affects considerably the
relaxation of the simulation cell. Previous studies of the silicon vacancy using supercells ranging from 32 to 216 atoms with various Brillouin zone sampling techniques~\cite{ Puska98,  Kum01, Ala93,  WA91, ES00, Pu00,Me98, PROB03, Len03} show a broad spread of formation energies (from 2.6 to 4.6 eV) and structural relaxation with ($D_{2d}$, $C_{2v}$, $C_{3v}$, $T_{d}$) symmetries. This scattering in mainly due to various convergence problems like  basis set size, Brillouin zone sampling and supercell size. 
A recent work of Probert and Payne~\cite{PROB03} propose a systematic methodology for accurate calculations of defect structure in supercells. We proceed in a similar way in  using supercells with up to 512-atoms.

\subsubsection{Mesh cut-off and the basis set }
\label{sec:pao}

With the presence of a defect, the single-$\zeta$ basis (SZ), i.e., a
single orbital per occupied state, is unsatisfactory: both the
formation energy and reconstruction surrounding the vacancy are far
from experimental value. Moreover, we find that the lack of overlap
between the PAO induced discontinuities in the force, causing serious
problems to the application of ART. It is possible to address these   
problems by using the optimized  basis generated using  the
  procedure of Anglada {\it et al.}~\cite{Ang02} (SZ-optimized). 
Although much improved, the discontinuity in the force remains due to the presence of an underlying grid for the integration in real space which causes breaking of the translational symmetry.
While the energy is converged with a real-space grid of about 0.03 \AA\, (a mesh cut-off of
35-40 Ry), we find that we need a grid space of at most 0.02 \AA\,
(mesh cut-off of 50 Ry) to stabilize the force derivative; this
spacing is comparable to the typical atomic displacement during the
dynamics monitored by ART.

\subsubsection{Brillouin zone sampling }
\label{sec:BZ}

Because the defect is highly localized even in large supercells, we need to have fully converged forces and total energies for the system. In order to isolate Brillouin zone sampling effects, we study force and energy convergence with respect to the density of k-points.
We use a Monkhorst and Pack~\cite{Mon76} mesh  to sample the Brillouin zone. Uniform meshes of $q^3$ shifted by ($k_0,k_0,k_0$) are used,  with $q$ going from 1 to 4 and $k_0$ = 0 or 0.5. For $q$=1 no offsets are used ($k_0=0$)  to sample the $\Gamma$ point. Offsets of $k_0=0.5$  are used for $q=2,3,4$ for each cubic supercell size. 

Figure~\ref{fig:bz} displays the total energy per atom embedded in differently sized supercells with respect to q. The total energy is converged to 1meV/atom at a mesh of $3^3$ for the 64 supercell, this corresponds  to a density of 0.031~\AA$^{-1}$ calculated using our LDA lattice parameter 5.39 ~\AA. As discussed by Probert and Payne~\cite{PROB03} this  density is sufficient to converge the total energy for larger systems. It is equivalent to a mesh of $2^3$ for 216 and 512 cells with densities of  0.031~\AA$^{-1}$ and 0.023~\AA$^{-1}$ respectively.
Comparing the forces for unrelaxed vacancy using different meshes and supercells of 63, 215 and 511 atoms, we find that atomic forces converge faster than the total energy as a function of $k$-point sampling:  convergence to 1 meV/\AA\  per degree of freedom is reached with a $2^3$ mesh for the 63 cell and with the $\Gamma$ point for larger cells.

\subsubsection{Supercell calculations }

\begin{table*}
\caption{Converged  formation energies for  ($E_f$) for different supercells and k-points meshes     compared with  various {\it ab-initio} calculations. For each work we specify the DFT functionals used: local density approximation (LDA), screened-exchange LDA (sx-LDA) and conjugate gradient approximation (GGA) with the corresponding plane-wave (PW) energy cutoff when applicable. (TB-EF) denotes tight binding calculations with eigenvalue following method. 
In the $D_{2d}$ symmetry or the pairing mode, the distorted tetrahedron formed by atoms around the vacant site  has  two short bonds for atoms belonging to the same pair (atom-atom) distance and four long bonds between atoms from different pairs (pair-pair).  The resulting symmetries of the defect  are reported as well. Experimentally, Watkins {\it et al.}~\cite{Corbett64} measured a formation energy of 3.6 $\pm$ 0.5~eV and Dannefaer {\it et al.}~\cite{Danne86}
3.6$\pm$0.2~eV}
\label{tab:dist}
\begin{ruledtabular}
\begin{tabular}{lllllllll}
Authors  &Method &Cell-size  & k-points   &$E_f$ (eV)  &\multicolumn{2}{c}{Distances (\AA)}  &Symmetry  \\
 &&&&&pair - pair  &atom - atom   \\
\hline
This work       &LDA &63  &$3^3$            &3.6   &3.34  &2.79  &$D_{2d}$\\
&&&$3^3$&3.95&3.19 &3.19  &$T_d$\\
                &  &215 &$\Gamma$  &3.1           &3.34  &2.89 &$D_{2d}$\\
                 &  &&$2^3$   &3.36           &3.35  &2.9 &$D_{2d}$\\
                 &  &511 &$2^3$   &3.29           &3.35  &2.9 &$D_{2d}$\\
 \\
Lento and Nieminen~\cite{Len03} & LDA   &32 &($\frac{1}{4}$,$\frac{1}{4}$,$\frac{1}{4}$) & &3.608 &3.084  &$D_{2d}$\\  
		&PW (15 Ry)	&256 &($\frac{1}{4}$,$\frac{1}{4}$,$\frac{1}{4}$)                 &3.6 & 3.490 & 2.950 &$D_{2d}$\\  
							
					&sx-LDA &256 &($\frac{1}{4}$,$\frac{1}{4}$,$\frac{1}{4}$)  &3.515 & & &$D_{2d}$-outward\\  
\\

Probert {\it et al.}~\cite{PROB03} & GGA  &256 &$2^3$ &3.17 &3.72 &3.1 &$D_{2d}$ \\
&PW (12Ry)\\
\\  
Puska {\it et al.}~\cite{Puska98}  &LDA           & 63 &$\Gamma$  &2.86 &    &&$C_{2v}$ \\
                             &PW (15 Ry) &     & $2^3$     &3.41 &3.42 & 3.40 &$D_{2d}$ \\ 
               &            &215 &$\Gamma$ &3.27   &3.38  &2.89 &$D_{2d}$\\
 \\
Mercer {\it et al.}~\cite{Me98} & LDA  &127 &$2^3$ &3.63  &3.50 &2.89 &$D_{2d}$\\

                          & PW (12-20 Ry)      \\
\\
Kumeda {\it et al.}~\cite{Kum01} & LDA &63 & $\Gamma$&3.32 \\ 
                    &PW(12 Ry)          &215 &$\Gamma$ &3.902\\ 
						& GGA & 215  &$\Gamma$  &3.336\\
\\
Seong and Lewis~\cite{Lewis96} &LDA  &63 &$\Gamma$ &3.65    & 3.53  & 3.00 &$D_{2d}$\\ 
			               	&PW (10 Ry) &511 &$\Gamma$         &4.12   \\ 
			\\

Munro and Wales ~\cite{Mu99} &TB-EF &  63  &$\Gamma$ &3.73 \\
                     &      &215  &$\Gamma$ &3.902\\
\\
Antonelli {\it et al.}~\cite{AN98} &LDA &63  &$2^3$    &~3.49 &3.53 &3.01 &$D_{2d}$\\  
	              &PW(16 Ry)&&&                   &3.52& 3.40 &$D_{2d}$\\
\\
							
\end{tabular}
\end{ruledtabular}
\end{table*}
As reported by Puska et al.~\cite{Pu00}, supercell method has obvious drawback because of the interaction of the defect and its periodic replicas. If the defect-defect distance is not large enough, the electronic structure  of an isolated defect is distorted  because the deep levels in the band gap form energy bands with a finite dispersion. The size of the supercell restricts also the ionic relaxation. The relaxation pattern is truncated midway between the defect and its nearest image.

In order to identify the smallest cell-size acceptable for the
simulation of diffusion mechanisms with negligible size effects, we
study the structural relaxation and the formation energies with respect
to supercell size by   employing supercells as large as  511 ionic sites.

Formation energies are calculated for fully converged supercells with respect to the basis set and Brillouin zone sampling as detailed in sections~\ref{sec:pao},~\ref {sec:BZ}. Results for 63 and 215 atoms supercells are overestimated compared to the 511-cell results by 9 \% and 2\% respectively as shown in table~\ref{tab:dist} . This error on vacancy formation energy  for  small supercell  was previously reported  by Puska {\it et al.}~\cite{Puska98} as  due to the  interaction of the vacancy with its periodic images and reflects a repulsive component of the defect-defect interaction, this component tends to vanish for the 215 supercell.

Structural relaxation is also of a major importance, it informs us  about the correctness of the atomic forces. The 63-atom supercell shows a wrong relaxation pattern. Tetrahedral ($T_d$ ) symmetry is a metastable minimum irrespective of the density of k points used with a formation energy of 3.99 eV ( see the second line on in table~\ref{tab:dist}). Symmetry need to be broken artificially  to get a Jahn-Teller distortion leading to  a $D_{2d}$  symmetry with lower formation energy (3.59 eV).  This is misleading and shows that size effects  are important for supercells of 63 ionic sites and smaller. This size is insufficient to represent the system and the diffusion pattern correctly  as  discussed in more details in section \ref {sec:size}. 
 
For larger system,  size effects tend to have minor effects on the relaxation geometry. Already at 215 cells, $D_{2d}$ symmetry is obtained even for low density of  k-points ($\Gamma$ only). The formation energy using $\Gamma $ point  for the 215 supercell is underestimated by 6\% compared to the best converged results we got with the 511 supercell,  it tends to  converge to the  formation energy  of larger models relaxed with $\Gamma$-point only.

We conclude, in agreement with previous works~\cite{PROB03, Pu00, Puska98} that a supercell of 215-atoms with $\Gamma$-point  is the minimal size and sampling  we can use in order to ensure that both  energy and  forces are sufficiently  accurate to provide results of a quality at least equal to experimental precision.

\section{Details of the simulation}
\label{sec:detail}

We perform first-principles electronic  structure calculations based on the density functional theory within the local-density approximation (LDA)~\cite{Cep80, Per81} for the exchange-correlation energy.  The static zero-temperature calculations for energies and forces are
computed from SIESTA.  Forces on the atoms are computed using the
Hellman-Feynman theorem. The nuclear  positions are optimized by using a
conjugate gradient algorithm (CG) to minimize the total energy. Using the unit cell of two atoms and increasing the number of k points the equilibrium lattice constant of bulk Si converges to 5.39 \AA. The cell volume and shape are kept fixed during the simulation. The
crystalline 216 supercell is relaxed until the largest force component
is of about 0.002~{eV/\AA}.  One conjugate gradient (CG) step is
necessary to relax the structure with vanishingly small force
components. To create the starting configuration with a defect, one
atom is removed from the ideal crystalline cell. The obtained
structure contains a mono-vacancy characterized by a vacancy-vacancy
distance of 16.17~\AA, with tetrahedral symmetry ($T_{d}$). The
structure is then allowed to relax at constant volume.
All the events are generated starting from a fully relaxed 215-atom
unit-cell with a single vacancy displaying an expected Jahn-Teller
distortion:  the four atoms around
the vacancy are paired two by two with $D_{2d}$ symmetry, as described
in section \ref{sec:relax}. More than 120 events were started from
this geometry, each launched in a different random direction,
providing an extensive sampling of the energy landscape around the
vacancy. The results from this search confirm  {\it a posteriori} that
this is indeed the minimum-energy configuration for the isolated
vacancy.
Using the set of parameters detailed in this section, one event with
SIEST-A-RT requires about 700 force evaluations and takes about 3
days of CPU times on a single IBM Power 4 processor. 

\section{Results}
\label{sec:Res}

\subsection{Neutral vacancy relaxation}
\label{sec:relax}

The details of the relaxation around the vacancy depend on the local charge~\cite{WATK92, lanoo}. Since there is one electron per dangling bond for the neutral charge state
($V^0$), it is expected to undergo a Jahn-Teller
distortion~\cite{WATK92, lanoo} by forming two pairs out of the four
dangling bonds of the vacancy. This is what we see. During the
relaxation of the 215-atom  cell, the total energy of the system is
reduced by about 1.71~eV, due to the distortion of the lattice and the
relaxation of the atoms around the vacancy. This relaxation energy is
larger than the 0.36~eV as obtained from a study on 63-atom cell by
Seong et Lewis~\cite{Lewis96} and the 0.9~eV calculated on a HF cluster
calculation~\cite{KIM97} and  compared to values from a more recent study ~\cite{PROB03} 1.186 eV using GGA on a 256 supercell.

\begin{figure}
\centerline{\includegraphics[width=4.8cm]{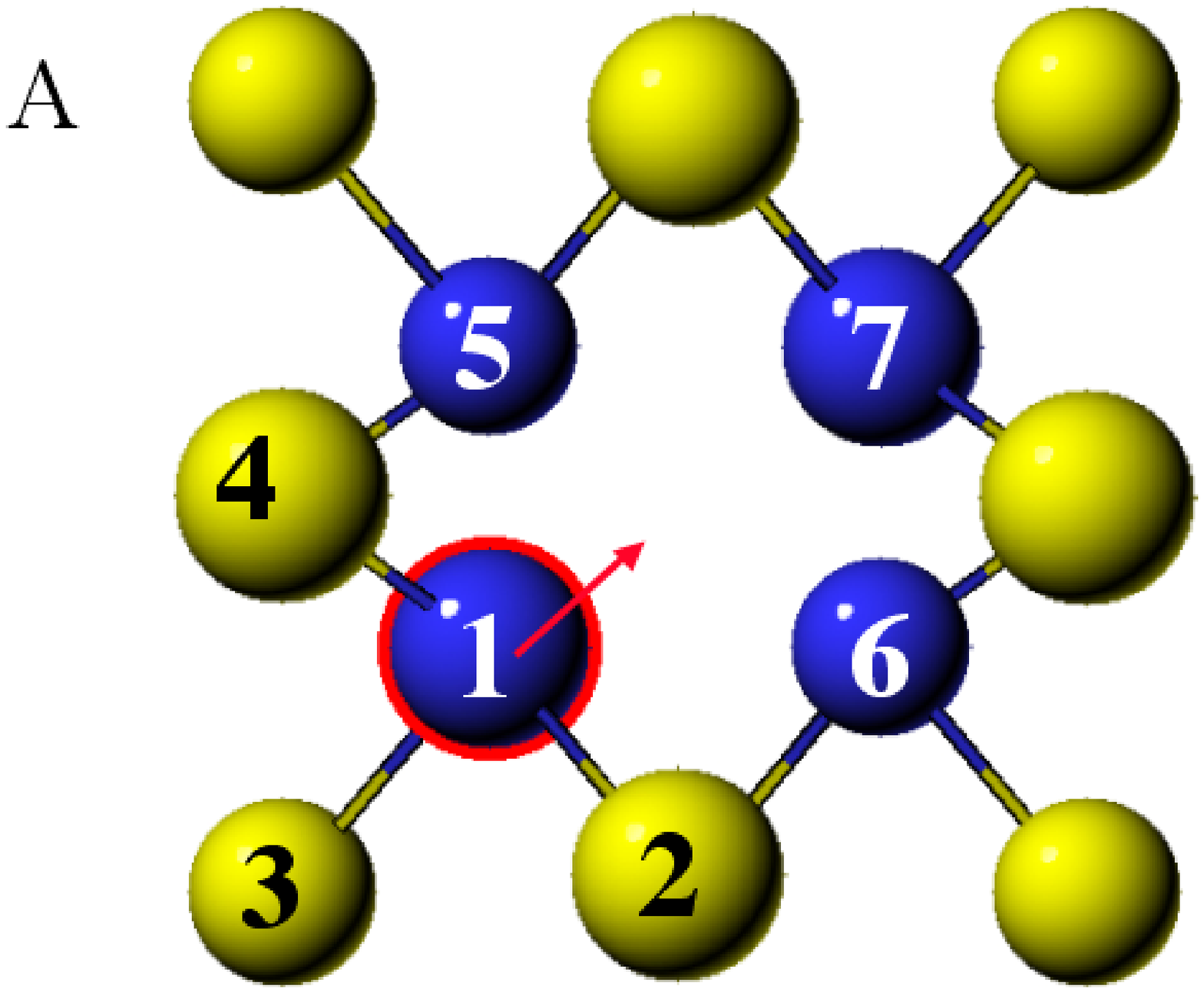} \includegraphics[width=4.8cm]{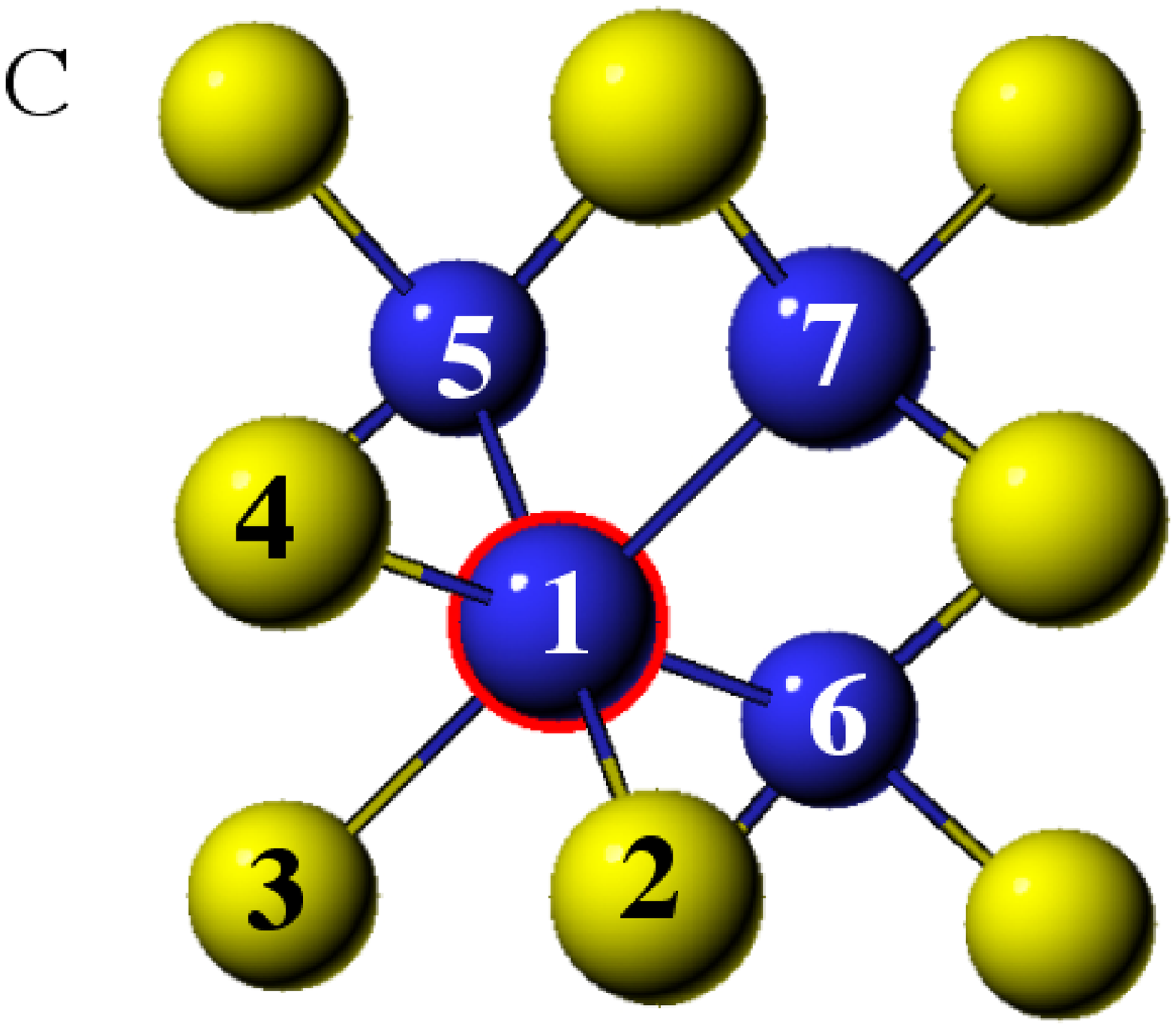}}
\caption{"(Color online)"  The vacant site is visible
  in the snapshots A and C.  A is the initial minimum and C is the
  ideal split vacancy site, the arrow shows the direction of the
  diffusion. }
\label{fig:snapshots}
\end{figure}

The relaxation can be described as follows.  The vacancy nearest
neighbors (NN) first move away from each other by pair along the $<110>$
axis~\cite {Lewis96}.  The net local displacement of 0.39~\AA\ per atom
result into  a tetragonal symmetry configuration ($D_{2d}$). The
distorted tetrahedron formed by the atoms has four long bonds of
3.35~\AA\ and two short bonds of 2.9~\AA\  as shown in
Fig.~\ref{fig:snapshots}(A). The formation of bonds between the two pairs
weakens the back bonds which are then significantly
stretched~\cite{AN98}, up to 2.49~\AA\ . The range of perturbation
introduced by the defect in the 215 supercell affects up to 5
surrounding shells: the average amplitude of relaxation falls
progessively from 0.39 \AA\ for the first shell to 0.074 \AA\ for the
fifth shell which is of the order of atomic vibrations at room
temperature.

The amplitude and the type of displacement compare very well with the
previous results using DFT~\cite{ Puska98,  Kum01, Ala93, WA91, ES00,
Pu00,Me98, PROB03}.  The LDA results appear to be at odds with recent
screened-exchange LDA calculations on partially relaxed neutral Si
vacancy, it indicates that the relaxation could be outward~\cite{Len03}.
More work remains to establish if LDA really fails here.  Our results on
the pairing mode are compared with some of these results  in table
\ref{tab:dist}.

The formation energy $E_f$ is calculated by taking into account the
distortion of the lattice, using the expression:
\beeq
E_f =E_{N-1} - \frac{N-1}{N}E_N
\eneq
Where N is the number of atoms and $E_N$ is the relaxed total energy
of the ideal structure. $E_{N-1}$ is the total energy of the distorted
system containing the defect.  Within SIESTA, the formation energy of
the mono-vacancy is 3.36~eV, which is in agreement with previous LDA
and experimental studies. On the simulation side, Kelly et
al.~\cite{Kelly92} calculated 3.5~eV, Bl\"ochl {\it et al.}~\cite{Blo93}
4.1~eV, Mercer {\it et al.}~\cite{Me98} 3.63~eV and Kumeda {\it et al.}~\cite
{Kum01} 3.32~eV.  Experimentally, Watkins {\it et al.}~\cite{Corbett64}
measured 3.6 $\pm$ 0.5~eV and Dannefaer {\it et al.}~\cite{Danne86}
3.6$\pm$0.2~eV.  A complete comparison in reported in table
\ref{tab:dist}.

The energy of 511 supercell, $E^{511}_{f}$, is found to
be 3.29~eV after a full convergence of the energy. Previous works by
Wang {\it et al.}~\cite{WA91} and Seong and Lewis~\cite {Lewis96} reported
 $E^{511}_{f}$= 4.12~eV and 4.10, respectively. A recent study by  Probert {\it et al.} give GGA results of   a fully converged calculation for a vacancy in BCC silicon lattice with 256 atoms, they obtain a formation energy of  3.17 eV.

\begin{figure}
\centerline{\includegraphics[width=8.cm]{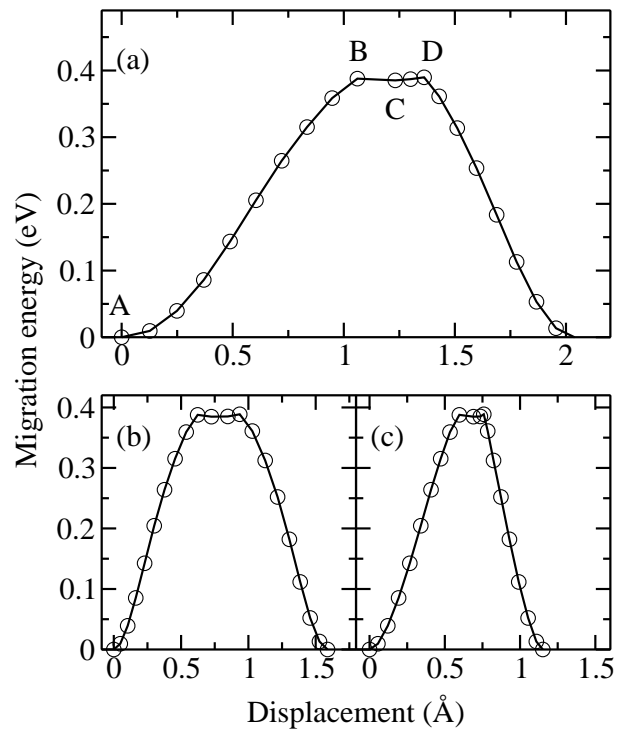}}

\caption{"(Color online)" Top:  (a) The total minimum-energy path for the simple diffusion. The saddle points
  identified in various ART events are indicated by the same labels (B,C and D) as in Table II.  The path is generated by previous
  knowledge of the initial, the saddle point and final states,  then
  the overall configuration is relaxed using the nudged-elastic-band
  method~\cite{UBER00}.  The contributions from different moving atoms are decomposed in two: (b) The path followed by the diffusing atom; (c) Diffusion of the other atoms around the defect. Here The assymetry is due to reconstruction.} 
\label{fig:diff-H}
\end{figure}

\subsection{Simple diffusion paths}
\label{sec:SDP}

Even a defect as simple as a neutral vacancy can be associated with
many activated mechanisms, some of which leading to migration:  
among the generated events, 60~\% are simple diffusion
mechanisms, while 15~\% are reorientation and the remaining are more
complex events involving a pair or more of active atoms.

\subsubsection{Nearest neighbor diffusion}

\begin{table}
\caption{Interatomic distances along the minimum energy path for a simple diffusion. The positions correspond to the labels in  Fig.~\ref{fig:diff-H}(a). A correcponds to the initial minimum with a relaxed vacancy:  the
  active atom (1) is bonded to four neighbors, B,C and D are transition states. During the migration  
  bonds are stretched  progressively until all the distances becomes equal at the
  configuration C.}
\label{tab:ABCD}
\begin{ruledtabular}
\begin{tabular} {p{2cm}p{1.cm}p{1.cm}p{1.cm}p{1.cm}p{1.cm}p{1.cm}p{1.cm}}

Distance(\AA) &$d_{1-2}$ &$d_{1-3}$ &$d_{1-4}$ &$d_{1-5}$ &$d_{1-6}$ &$d_{1-7}$\\
\hline
\bf{A} &2.49 &2.39 &2.39 &2.89 & 3.34 &3.34\\
\bf{B}  &2.67 &2.67 &2.65 &2.65 &2.76 &2.73\\
\bf{C} &2.68 &2.68 &2.68 &2.68 &2.68 &2.68\\
\bf{D} &2.68 &2.87 &2.61 &2.69 &2.61 &2.69\\ \\
\end{tabular}
\end{ruledtabular}
\end{table}

The simplest migration process involves hopping of the vacancy to a nearest-neighbor site as shown in Fig.~\ref{fig:diff-H}(a): a nearest-neighbor of the vacancy moves along the $<111>$ direction toward the vacant site (shown in  Fig.~\ref{fig:snapshots}A) and passes the metastable split-vacancy site before falling in the tetrahedral site previously formed by the vacancy. The energy barrier for this mechanism is $0.40\pm0.02$ eV. The error estimate comes from the convergence criterion  to the saddle point as explained in section \ref{sec:ART}. This result for the migration energy is in good agreement with the experimental findings of Watkins et al.~\cite{WATK92} who measured $0.45\pm$0.04~eV. In a work similar to this one, but using the CPMD code with GGA (BLYP) functional, Kumeda {\it et al.}~\cite{Kum01}found a barrier of 0.58~eV.

The diffusion path is reconstructed in detail using the nudged-elastic band method\cite{UBER00} and relaxed until the total force is of 0.2 eV/\AA. This path  shows an unusually long transition plateau ---ranging from $d_r$=0.7 to 1.2 \AA--- with a slight energy minimum at the split-vacancy site. At this point, the 3 back bonds are stretched and the diffusing atom forms weak bonds (length 2.685\AA) with the surrounding atoms of the six membered ring. From this site, a small displacement of 0.11 \AA\ is sufficient to overcome the barrier of 0.04~eV and complete the jump to the new minimum (see table \ref{tab:ABCD}). This value is of the order of the precision of the method;  in addition,  at room temperature  this minimum would be smeared out by thermal vibrations. Because of a lack of a better approximation, we use Vineyard's quasiharmonic approximations~\cite{Vin57} to get a rough estimate of the attempt frequency (see section \ref{sec:diffusion}),   approximating the transition plateau by a  single barrier.

 Although the total diffusion path in Fig.~\ref{fig:diff-H}(a) is asymmetric, after separating the contribution coming  from different moving atoms, we find that diffusion path  is symmetric for the diffusing atom as shown in figure (Fig.~\ref{fig:diff-H}(b)). The asymmetry in the path  shown in Fig.~\ref{fig:diff-H}(c) 
 is due to the reconstruction: with four atoms reconstructing two-by-two, there are three possible orientations. Here, the orientation is different for the initial and final state. This should average out as the vacancy moves around.

 A comparison of the formation energies using the
Stillinger-Weber~\cite{Maroud93} and the MEAM~\cite{Baskes} potentials
for the vacancy and the split vacancy gives a migration energy of 0.43
and 0.37 eV, respectively.  Although the formation energies for the
vacancy are not well represented by empirical potentials the energy
difference is in agreement with experiment and with our results.  The
minimum-energy path, however, is not well described by these potential
and the local minima identified by Maroudas {\it et
  al.}~\cite{Maroud93} during the migration are absent in our case.

\subsubsection{Reorientation process}
\label{sec:reo}

In addition to finding diffusion mechanisms associated with the
vacancy, we also identify the trajectories responsible for atomic
rearrangement in the reconstructed configuration. As there are three
pairing possibilities for the $D_{2d}$ symmetry, it is possible for
the configuration to hop from one of these states to either of the two
others. 

\begin{figure}
\leftline{\includegraphics[width=10cm]{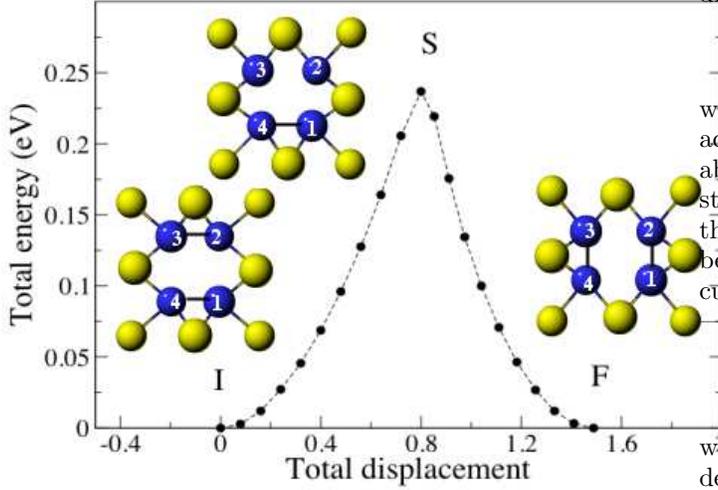}}
\caption{"(Color on line)" The reorientation path passing over the saddle point
  configuration. (I$\to$S$\to$F), the tetragonal axis is fixed
  throughout this process. At the saddle point the local atomic
  coordinates of the distorted lattice shows a trigonal symmetry. The
  pairing mode changes to one of the equivalent modes.}
\label{fig:reor}
\end{figure}

SIEST-A-RT generates easily this reorientation of the paired atoms. 
As illustrated in Fig.~\ref{fig:reor}, starting with the reconstructed pairs
formed initially --- 1-4 and 2-3---, atom ``3'' moves away form
its neighbor along the $<111>$ direction, breaks the weak bond with
"2" and reaches the transition state, which is in the
trigonal symmetry ($C_{3v}$). The relaxation continues until
new bonds form between the pairs 1-2 and 3-4.  The activation energy
for the reorientation of the neutral vacancy between two tetragonal
orientations is 0.2~eV. Roberson {\it et al.} \cite{ROBER94} in a HF
calculation scheme performed calculation on small clusters (44 silicon
atoms) and found 0.37~eV; another cluster calculation, by \"O\u{g}\"ut
{\it et al.}~\cite{KIM97} found 0.32~eV. 

The deformation to the $C_{3v}$ can be observed experimentally. The experiment
consists of applying a stress at high temperature in the dark, then
cooling down to about 20K, relieving the stress, illuminating the
sample and monitoring the resulting vacancy species. The barrier
measured from this alignment experiment is at 0.23~eV, in excellent
agreement with our calculations~\cite{WATK92}.

Our simulations also show that the pairing mode can change {\it
  during} the migration of the vacancy.  For the simple diffusion plus
reorientation, we find an overall energy barrier of 0.47 eV and a
total displacement of 0.68~\AA\  at the saddle point.

\subsubsection{Diffusion constants and diffusion rate}
\label{sec:diffusion}

The diffusion is a two step process, consisting of the creation and
the subsequent migration of a vacancy. Assuming that both processes
are thermally activated, the activation energy $E_a$ for diffusion is
the sum of the formation energy $E_f$ and the migration energy
$E_m$. By collecting our data from the previous sections we get an
activation energy of $E_a$ = 3.5 eV, which is in good agreement with
 $E_a=4.14$ eV recently 
measured by Bracht {\it et al.}~\cite{Bracht98}, and with
the activation enthalpy for self-diffusion $H_V$ = 4.07 $\pm$ 0.2 eV
obtained by Tang {\it et al.}~\cite{Tang97} using tight-binding MD.

It is well known that there is a strong competition between the
interstitial and the vacancy-mediated mechanisms of
self-diffusion. These two contributions were separated in a
metal-diffusion experiment~\cite{Bracht98} and the vacancy
contribution to the self-diffusion was measured to be
\beeq
C_{V}^{eq}(T)d_v(T) = 0.92 exp\frac{(-4.14 eV)}{k_BT} cm^2s^{-1}
\eneq
Other experimental data~\cite{Tan85} shows that 
\beeq
C_{V}^{eq}(T)d_v(T) = 0.6 exp\frac{(-4.03 eV)}{k_BT} cm^2s^{-1}
\eneq
Both data sets predict a self-diffusion dominated by vacancies at low
temperature, where $C_{V}^{eq}(T)$ is the concentration of  vacancies
at a certain temperature.

Experiments under thermal equilibrium concentrations of native point
defects have shown that the Si self-diffusion coefficient
$D_{Si}^{eq}$ exhibits Arrhenius behavior,
\beeq
D_{Si}^{eq}(T) = D_0e^{-H_a/k_bT},
\eneq
where $H_a$ is a single activation enthalpy, $D_0$ is the prefactor in
a certain temperature range, $T$ denotes the absolute temperature, and
$k_B$ is the Boltzman's constant. Under the assumption that the
diffusion proceeds through discrete jumps of equal length $r$ ---the
distance between nearest neighbor equilibrium sites, in diamond cubic
lattices $r=a\sqrt{3}/4$, where $a$ is the lattice parameter --- 
the prefactor is written as~\cite{Lopez88}
\beeq
D_0 = {\frac{z}{6} \Gamma_0 r^2}
\eneq
with ${\Gamma}_0$ the jump, or attempt, frequency of the point defect from one
equilibrium site to another, z denotes the number
of neighbor sites, in the diamond lattice a vacancy can jump to one of the  four neighbors.
In the harmonic approximation, this attempt frequency is defined as~\cite{Vin57, Rob01}
\beeq
\Gamma_0 =   \frac{ \displaystyle \prod_{i=1}^{3N}
    \nu_i^{(i)}} {\displaystyle \prod_{i=1}^{3N-1} \nu_i^{(s)}}
\eneq
where $\nu_i^{(i)}$ and $\nu_i^{(S)}$ are the normal mode frequencies at the minimum and  the saddle point respectively and the product does not include the imaginary frequency at the saddle point. The eigenvalues are computed by diagonalizing the Hessian, obtained by numerical derivation with a step of 0.01 \AA.

Accordingly, the migration entropy defined as the entropy difference
between the saddle point and the equilibrium point configuration at low temperature can be written as
\beeq
\Delta S = S_V^m =k_B \ln \left[ \frac{ \displaystyle \prod_{i=1}^{3N-1}
    \nu_i^{(i)}} {\displaystyle  \prod_{i=1}^{3N-1} \nu_i^{(s)}}\right]
\eneq
The phonon frequency corresponding the direction of the jump is removed from the  numerator. In addition
the self diffusion entropy evaluated experimentally is the sum of the
diffusion entropy and the formation entropy $S^{SD}_V=S_V^f + S_V^m$.
Using the experimental $S^{SD}_V=5.5k_B$ by Bracht {\it et
  al.}~\cite{Bracht98} and a $S_v^f = (5\pm2)k_B$ as calculated by
Bl\"ochl {\it et al.}~\cite{Blo93} and Dobson {\it et
  al.}~\cite{Vech89}, migration entropy is expected to lie within
1$k_B $ and 2.5$k_B$,  our calculated data $\Delta S =
3.036 k_B$.

The corresponding attempt frequency $\Gamma_0$ = $8.65\times10^{12}
s^{-1}$ is similar to the result of Maroudas et
al.~\cite{Maroud93} ($1.0539\times10^{13} s^{-1}$), obtained with the
Stillinger-Weber potential. The 
corresponding diffusion prefactor is $D_0 =3.141\times10^{-3} cm^2s^{-1}$. Our diffusion constant can be compared with results from  Maroudas {\it et al.}~\cite{Maroud93} by taking into account the z factor that had been omitted, the  resulting prefactor   $D_0 =3.88\times10^{-3} cm^2s^{-1}$ is in agreement with our results. Tang {\it et al.}~\cite{Tang97,Colom01} report  $D_0 =1.18\times10^{-4}cm^2s^{-1}$ using  TB-MD and thermodynamical integration, their migration energy ($E_m$) is of 0.1 eV.
	
Our numerical estimate of the diffusion rate $\Gamma=\Gamma_0
e^{E_m/k_BT}$ at room temperature gives a characteristic time for diffusion ($\tau = 1/\Gamma$ ) of the order of microseconds. 
This timescale is not directly accessible to {\it ab-initio} molecular dynamics, so
simulations must generally be run at temperatures close to the
melting point. With SIEST-A-RT, we can study the activated mechanisms
from the local energy minimum, allowing us to identify accurately
subtle mechanisms such as that associated with the split-vacancy site.

\subsection{Complex diffusion paths }

The SIEST-A-RT sampling of the energy landscape around the vacancy
also identifies a number of high-energy barrier events which involve a
complex atomistic rearrangements. Two types of such events occur. The
first one can be described as a diffusion event mediated by bond
exchange. In this case, the overall result is a simple vacancy hop but
with a more complex transition state and  a barrier of 2.4 eV. 

\begin{figure}
\centerline{\includegraphics[width=9.5cm]{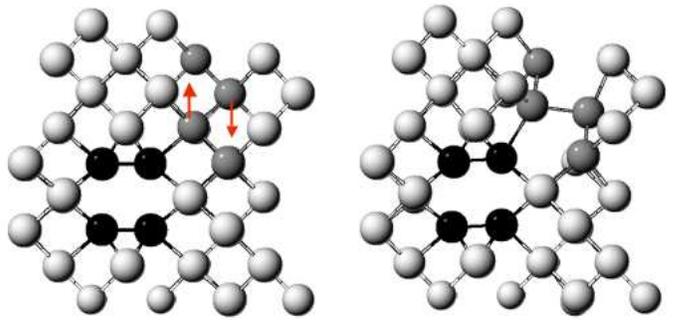}}
\caption{"(Color on line)"  A {\it spectator} WWW process involving atoms far from the vacant site. Left
  panel: atoms in grey are the atoms involved in the exchange,  the
  direction of the exchange is illustrated by the arrows. Right panel:
  the final configuration is a stable minimum by which a disorder in
  the crystalline lattice is introduced.}
\label{fig:W3}
\end{figure}

The second type of events involves what has been dubbed {\it spectator}
  events~\cite{Kum01}  (see Fig.~\ref{fig:W3}) where the diamond network
surrounding the vacancy is alterated. These spectator events all correspond basically to the bond-switching mechanisms proposed many years ago by Wooten, Winer and
Weaire as mechanism for amorphization~\cite{WWW85}. In recent years,
the WWW mechanism has also been identified as the bond defect~\cite
{CARG98} and the four fold coordinated defect (FFCD)~\cite{Goe02}.

Two types of {\it spectator} events are identified.
In one  of these events, two of the direct
neighbors of the vacancy are pushed toward each other, leading to the
saturation of the dangling bonds; as a consequence one of the long
covalent bonds between the NN of the vacancy transform to a real bond
in a form very similar to the bond defect complex identified by
Marqu\'es et al~\cite{AM01}.  The lowest barrier we found is 2.63 eV
for a process involving a direct neighbor of the vacancy (one of the
bonds is already weak), the final configuration is 2.16 eV higher than
the initial minimum. Kumeda {\it et al.} found the single {\it spectator} WWW barrier of
1.719 eV using LDA and 1.867 eV using BLYP functional~\cite{Kum01}.

The second type, atoms far from the vacant site are involved. The effect of the vacancy on the WWW event barrier is less strong for complex events located far from the vacancy without involving a direct neighbor. As the lattice distortion increases, so does
the energy barrier. For this mechanism, for example, the barrier ranges
from 3.31-3.8 eV, a value which can be compared to the I-V
recombination process identified to be responsible for the
amorphization of crystalline silicon of Marqu\'es et al~\cite{AM01}
the barrier is estimated to be between 4.58 and 4.75 eV using {\it ab
  initio} calculation~\cite {CARG98}. One can estimate the effect of the vacancy to lower the barrier by about 1.27 eV. Since the barrier is very high,
this kind of events is unlikely to play any significant role in
crystals at room temperature.

\subsection{Size effects}
\label{sec:size}

As discussed, previously, it is necessary to use a large simulation
cell in order to minimize the interaction between the images of the
vacancy. This is particularly noticeable in a 63-atom supercell, even
though the defect-defect separation is larger than 4 inter-atomic
distances. It is this interaction that cause the unrealistic
relaxation pattern identified in previous studies~\cite{Me98, Ala93},
allowing the vacancy to relax both with a $ T_{2d} $ and a $ D_{2d}$
symmetries. 

Relaxing a 63-atom supercell with SIESTA and the parameters described
above, we find that two stable minima:  the $D_{2d}$ (tetragonal)
relaxation described previously as well as a $T_{d}$ (tetrahedral)
configuration, a symmetry which is {\it unstable} in cells of 215 and 511 atoms. Within the later symmetry, all atoms relax inward,
conserving the tetrahedral symmetry of the system at equal distance of
3.4~\AA.  The $T_{2d} $ structure can be reached with a single jump
from the $D_{2d}$ site during a reorientation mechanism similar to
that described in section \ref{sec:reo}, crossing a small barrier of
of 94 meV with an overall displacement of 0.64 \AA. The energy
difference between the two symmetries is estimated to be 83 meV compared
to 20~meV as obtained by Mercer {\it et al.} in a simulation also on a
63-atom cell~\cite{Me98}. We obtain qualitatively similar results with $2^3$ k-points mesh.

The interaction between images also affects the barrier for the
simple diffusion and the reorientation, which is underestimated by 50 \%
compared to the 215 cell results. Interestingly, this underestimation
is compensated by a higher formation energy, leading to very similar
activation energies for the two size systems. 

The underestimation of the diffusion barrier in the 63-atom cell
makes it easier for the vacancy to jump to the second-neighbor
position. Although we saw many such events in the small cell, this
mechanism was not generated in the 215-atom box; there does not
appear to be any direct path for second-neighbor diffusion in
silicon.
\section{Discussion and Conclusions}
\label{sec:discussion}

In this paper, we report on the study of the activated dynamics
associated with a neutral vacancy in bulk crystalline
silicon. Following the work of Munro,
Kumeda and Wales~\cite{Mu99,Kum01}, we couple an {\it ab initio} code,
SIESTA, with the activation-relaxation technique, ART nouveau, in
order to sample the various activated mechanisms taking place around
the vacancy accurately and efficiently. Defining moves directly in the
energy landscape, it is as easy to generate the diffusion trajectory
responsible for the high-energy bond-switching mechanisms than it is
to find the path associated with the jump reorientation of the
reconstructed state ( $D_{2d}\to D_{2d}$).

Simulating more than 120 activated trajectories, we find a number of
different diffusion mechanisms associated with the vacancy. In
particular, we recover the basic vacancy hopping diffusion mechanism
and show that the path associated with it includes a very metastable
state at the split-vacancy site. Moreover, we can reproduce without
difficulty the jump between the symmetrically equivalent
reconstructed states. We also generate a number of higher-energy
diffusion mechanisms which reminiscent of those found in amorphous
silicon as well as at the crystalline-amorphous interface. 

Because we obtain, with good accuracy, the transition state
associated with each of these mechanisms, we also compute the
transition rate in the harmonic approximation. The result obtained is
in good agreement with experiment as well as previous
high-temperature TB-MD simulation~\cite{Tang97}.
 
Overall,  various barrier and relaxation energies computed here are
in agreement with results obtained previously both experimentally and
numerically.  Using a large simulation cell and an extensive sampling
of the energy landscape around the vacancy, we could also show that
some of the previously obtained mechanisms were artifact due to the
small size of the cell used.  These results establish the validity of
the SIEST-A-RT approach as a unique tool for the study of the
diffusion and relaxation mechanisms of defects in crystalline
materials.  This method can be easily extended to study semiconductor
compounds, adatom diffusion, interlayer diffusion, all systems with a
landscape too complicated for high-temperature molecular dynamics.
Even in system as simple as a vacancy in {\it c}-Si, we found a number
of activated mechanism associated with non-trivial transition
paths. The situation promises to be more complex in binary
semiconductors, such as GaAs and GaN, where the bonding character and
chemical composition are much more complex than for Si.

\section{Acknowledgments}
This work is funded in part by NSERC and NATEQ. NM is a Cottrell
Scholar of the Research Corporation. Most of the simulations were run on the computers of the R\'eseau qu\'eb\'ecois de calcul de haute performance (RQCHP) whose support is gratefully acknowledged.


\begin{thebibliography}{59}
\expandafter\ifx\csname natexlab\endcsname\relax\def\natexlab#1{#1}\fi
\expandafter\ifx\csname bibnamefont\endcsname\relax
  \def\bibnamefont#1{#1}\fi
\expandafter\ifx\csname bibfnamefont\endcsname\relax
  \def\bibfnamefont#1{#1}\fi
\expandafter\ifx\csname citenamefont\endcsname\relax
  \def\citenamefont#1{#1}\fi
\expandafter\ifx\csname url\endcsname\relax
  \def\url#1{\texttt{#1}}\fi
\expandafter\ifx\csname urlprefix\endcsname\relax\def\urlprefix{URL }\fi
\providecommand{\bibinfo}[2]{#2}
\providecommand{\eprint}[2][]{\url{#2}}

\bibitem[{\citenamefont{Ural et~al.}(1999)\citenamefont{Ural, Griffin, and
  Plummer}}]{URAL99}
\bibinfo{author}{\bibfnamefont{A.}~\bibnamefont{Ural}},
  \bibinfo{author}{\bibfnamefont{P.~B.} \bibnamefont{Griffin}},
  \bibnamefont{and} \bibinfo{author}{\bibfnamefont{J.~D.}
  \bibnamefont{Plummer}}, \bibinfo{journal}{Phys.\ Rev.\ Lett.}
  \textbf{\bibinfo{volume}{83}}, \bibinfo{pages}{3454} (\bibinfo{year}{1999}).

\bibitem[{\citenamefont{Lagna and Coffa}(1999)}]{Ma99}
\bibinfo{author}{\bibfnamefont{A.~L.} \bibnamefont{Lagna}} \bibnamefont{and}
  \bibinfo{author}{\bibfnamefont{S.}~\bibnamefont{Coffa}},
  \bibinfo{journal}{Phys.\ Rev.\ Lett.} \textbf{\bibinfo{volume}{82}},
  \bibinfo{pages}{1720} (\bibinfo{year}{1999}).

\bibitem[{\citenamefont{Watkins}(1992)}]{WATK92}
\bibinfo{author}{\bibfnamefont{G.~D.} \bibnamefont{Watkins}},
  \emph{\bibinfo{title}{Deep Centers in semiconductors, edited by S. T.
  Pantelides}} (\bibinfo{publisher}{Gordon and Breach}, \bibinfo{address}{New
  York}, \bibinfo{year}{1992}), chap.~\bibinfo{chapter}{3}.

\bibitem[{\citenamefont{Watkins and Corbett}(1964)}]{Corbett64}
\bibinfo{author}{\bibfnamefont{G.}~\bibnamefont{Watkins}} \bibnamefont{and}
  \bibinfo{author}{\bibfnamefont{J.}~\bibnamefont{Corbett}},
  \bibinfo{journal}{Phys. Rev.} \textbf{\bibinfo{volume}{134}},
  \bibinfo{pages}{A1389} (\bibinfo{year}{1964}).

\bibitem[{\citenamefont{Dannefaer et~al.}(1986)\citenamefont{Dannefaer,
  Mascher, and Kerr}}]{Danne86}
\bibinfo{author}{\bibfnamefont{S.}~\bibnamefont{Dannefaer}},
  \bibinfo{author}{\bibfnamefont{P.}~\bibnamefont{Mascher}}, \bibnamefont{and}
  \bibinfo{author}{\bibfnamefont{D.}~\bibnamefont{Kerr}},
  \bibinfo{journal}{Phys.\ Rev.\ Lett.} \textbf{\bibinfo{volume}{56}},
  \bibinfo{pages}{2195} (\bibinfo{year}{1986}).

\bibitem[{\citenamefont{Ghaisas}(1991)}]{Ghaisas91}
\bibinfo{author}{\bibfnamefont{S.~V.} \bibnamefont{Ghaisas}},
  \bibinfo{journal}{Phys.\ Rev.\ B} \textbf{\bibinfo{volume}{43}},
  \bibinfo{pages}{1808} (\bibinfo{year}{1991}).

\bibitem[{\citenamefont{\"O\u{g}\"ut and Chelikowsky}(1999)}]{Cheli99}
\bibinfo{author}{\bibfnamefont{S.}~\bibnamefont{\"O\u{g}\"ut}}
  \bibnamefont{and} \bibinfo{author}{\bibfnamefont{J.~R.}
  \bibnamefont{Chelikowsky}}, \bibinfo{journal}{Phys.\ Rev.\ Lett.}
  \textbf{\bibinfo{volume}{83}}, \bibinfo{pages}{3852} (\bibinfo{year}{1999}).

\bibitem[{\citenamefont{J\"a\"askel\"ainen.
  et~al.}(2001)\citenamefont{J\"a\"askel\"ainen., Colombo, and
  Nieminen}}]{Colom01}
\bibinfo{author}{\bibfnamefont{A.}~\bibnamefont{J\"a\"askel\"ainen.}},
  \bibinfo{author}{\bibfnamefont{L.}~\bibnamefont{Colombo}}, \bibnamefont{and}
  \bibinfo{author}{\bibfnamefont{R.}~\bibnamefont{Nieminen}},
  \bibinfo{journal}{Phys.\ Rev.\ B} \textbf{\bibinfo{volume}{64}},
  \bibinfo{pages}{233203} (\bibinfo{year}{2001}).

\bibitem[{\citenamefont{Dobson et~al.}(1989)\citenamefont{Dobson, Wager, and
  Van~Vechten}}]{Vech89}
\bibinfo{author}{\bibfnamefont{T.~W.} \bibnamefont{Dobson}},
  \bibinfo{author}{\bibfnamefont{J.~F.} \bibnamefont{Wager}}, \bibnamefont{and}
  \bibinfo{author}{\bibfnamefont{J.~A.} \bibnamefont{Van~Vechten}},
  \bibinfo{journal}{Phys.\ Rev.\ B} \textbf{\bibinfo{volume}{40}},
  \bibinfo{pages}{2962} (\bibinfo{year}{1989}).

\bibitem[{\citenamefont{Seong and Lewis}(1996)}]{Lewis96}
\bibinfo{author}{\bibfnamefont{H.}~\bibnamefont{Seong}} \bibnamefont{and}
  \bibinfo{author}{\bibfnamefont{L.~J.} \bibnamefont{Lewis}},
  \bibinfo{journal}{Phys.\ Rev.\ B} \textbf{\bibinfo{volume}{53}},
  \bibinfo{pages}{9791} (\bibinfo{year}{1996}).

\bibitem[{\citenamefont{Puska et~al.}(1998)\citenamefont{Puska, P\"oykk\"o,
  Pesola, and Nieminen}}]{Puska98}
\bibinfo{author}{\bibfnamefont{M.~J.} \bibnamefont{Puska}},
  \bibinfo{author}{\bibfnamefont{S.}~\bibnamefont{P\"oykk\"o}},
  \bibinfo{author}{\bibfnamefont{M.}~\bibnamefont{Pesola}}, \bibnamefont{and}
  \bibinfo{author}{\bibfnamefont{R.~M.} \bibnamefont{Nieminen}},
  \bibinfo{journal}{Phys.\ Rev.\ B} \textbf{\bibinfo{volume}{58}},
  \bibinfo{pages}{1318} (\bibinfo{year}{1998}).

\bibitem[{\citenamefont{Antonelli et~al.}(1998)\citenamefont{Antonelli,
  Kaxiras, and Chadi}}]{AN98}
\bibinfo{author}{\bibfnamefont{A.}~\bibnamefont{Antonelli}},
  \bibinfo{author}{\bibfnamefont{E.}~\bibnamefont{Kaxiras}}, \bibnamefont{and}
  \bibinfo{author}{\bibfnamefont{D.~J.} \bibnamefont{Chadi}},
  \bibinfo{journal}{Phys.\ Rev.\ Lett.} \textbf{\bibinfo{volume}{81}},
  \bibinfo{pages}{2088} (\bibinfo{year}{1998}).

\bibitem[{\citenamefont{Bar-Yam and Joannnopoulos}(1984)}]{Yam84}
\bibinfo{author}{\bibfnamefont{Y.}~\bibnamefont{Bar-Yam}} \bibnamefont{and}
  \bibinfo{author}{\bibfnamefont{J.~D.} \bibnamefont{Joannnopoulos}},
  \bibinfo{journal}{Phys.\ Rev.\ B} \textbf{\bibinfo{volume}{30}},
  \bibinfo{pages}{2216} (\bibinfo{year}{1984}).

\bibitem[{\citenamefont{Kumeda et~al.}(2001)\citenamefont{Kumeda, Wales, and
  Munro}}]{Kum01}
\bibinfo{author}{\bibfnamefont{Y.}~\bibnamefont{Kumeda}},
  \bibinfo{author}{\bibfnamefont{D.~J.} \bibnamefont{Wales}}, \bibnamefont{and}
  \bibinfo{author}{\bibfnamefont{L.~J.} \bibnamefont{Munro}},
  \bibinfo{journal}{Chem.\ Phys.\ Lett.} \textbf{\bibinfo{volume}{341}},
  \bibinfo{pages}{185} (\bibinfo{year}{2001}).

\bibitem[{\citenamefont{Virkkunen et~al.}(1993)\citenamefont{Virkkunen,
  Alatalo, Puska, and Nieminen}}]{Ala93}
\bibinfo{author}{\bibfnamefont{R.}~\bibnamefont{Virkkunen}},
  \bibinfo{author}{\bibfnamefont{M.}~\bibnamefont{Alatalo}},
  \bibinfo{author}{\bibfnamefont{M.}~\bibnamefont{Puska}}, \bibnamefont{and}
  \bibinfo{author}{\bibfnamefont{R.~M.} \bibnamefont{Nieminen}},
  \bibinfo{journal}{Comp.\ Mat.\ Sc.} \textbf{\bibinfo{volume}{1}},
  \bibinfo{pages}{151} (\bibinfo{year}{1993}).

\bibitem[{\citenamefont{Goedecker et~al.}(2002)\citenamefont{Goedecker,
  Billard, and Deutch}}]{Goe02}
\bibinfo{author}{\bibfnamefont{S.}~\bibnamefont{Goedecker}},
  \bibinfo{author}{\bibfnamefont{L.}~\bibnamefont{Billard}}, \bibnamefont{and}
  \bibinfo{author}{\bibfnamefont{T.}~\bibnamefont{Deutch}},
  \bibinfo{journal}{Phys.\ Rev.\ Lett.} \textbf{\bibinfo{volume}{88}},
  \bibinfo{pages}{235501} (\bibinfo{year}{2002}).

\bibitem[{\citenamefont{Marques et~al.}(2001)\citenamefont{Marques, Pelaz,
  Hernandez, Barbolla, and Gilmer}}]{AM01}
\bibinfo{author}{\bibfnamefont{L.}~\bibnamefont{Marques}},
  \bibinfo{author}{\bibfnamefont{L.}~\bibnamefont{Pelaz}},
  \bibinfo{author}{\bibfnamefont{J.}~\bibnamefont{Hernandez}},
  \bibinfo{author}{\bibfnamefont{J.}~\bibnamefont{Barbolla}}, \bibnamefont{and}
  \bibinfo{author}{\bibfnamefont{G.~H.} \bibnamefont{Gilmer}},
  \bibinfo{journal}{Phys.\ Rev.\ B} \textbf{\bibinfo{volume}{64}},
  \bibinfo{pages}{045214} (\bibinfo{year}{2001}).

\bibitem[{\citenamefont{Kim et~al.}(2000)\citenamefont{Kim, Kirchhoff, Watkins,
  and Khan}}]{Kim00}
\bibinfo{author}{\bibfnamefont{J.}~\bibnamefont{Kim}},
  \bibinfo{author}{\bibfnamefont{F.}~\bibnamefont{Kirchhoff}},
  \bibinfo{author}{\bibfnamefont{J.}~\bibnamefont{Watkins}}, \bibnamefont{and}
  \bibinfo{author}{\bibfnamefont{F.~S.} \bibnamefont{Khan}},
  \bibinfo{journal}{Phys.\ Rev.\ Lett.} \textbf{\bibinfo{volume}{84}},
  \bibinfo{pages}{503} (\bibinfo{year}{2000}).

\bibitem[{\citenamefont{Wang et~al.}(1991)\citenamefont{Wang, Chan, and
  Ho}}]{WA91}
\bibinfo{author}{\bibfnamefont{C.~Z.} \bibnamefont{Wang}},
  \bibinfo{author}{\bibfnamefont{C.~T.} \bibnamefont{Chan}}, \bibnamefont{and}
  \bibinfo{author}{\bibfnamefont{K.~M.} \bibnamefont{Ho}},
  \bibinfo{journal}{Phys.\ Rev.\ Lett.} \textbf{\bibinfo{volume}{66}},
  \bibinfo{pages}{189} (\bibinfo{year}{1991}).

\bibitem[{\citenamefont{Tang et~al.}(1997)\citenamefont{Tang, Colombo, Zhu, and
  de~la Rubia}}]{Tang97}
\bibinfo{author}{\bibfnamefont{M.}~\bibnamefont{Tang}},
  \bibinfo{author}{\bibfnamefont{L.}~\bibnamefont{Colombo}},
  \bibinfo{author}{\bibfnamefont{J.}~\bibnamefont{Zhu}}, \bibnamefont{and}
  \bibinfo{author}{\bibfnamefont{T.~D.} \bibnamefont{de~la Rubia}},
  \bibinfo{journal}{Phys.\ Rev.\ Lett.} \textbf{\bibinfo{volume}{55}},
  \bibinfo{pages}{14279} (\bibinfo{year}{1997}).

\bibitem[{\citenamefont{Leung et~al.}(1999)\citenamefont{Leung, Needs, and
  Rajogopal}}]{Le99}
\bibinfo{author}{\bibfnamefont{W.~K.} \bibnamefont{Leung}},
  \bibinfo{author}{\bibfnamefont{R.~J.} \bibnamefont{Needs}}, \bibnamefont{and}
  \bibinfo{author}{\bibfnamefont{G.}~\bibnamefont{Rajogopal}},
  \bibinfo{journal}{Phys.\ Rev.\ Lett.} \textbf{\bibinfo{volume}{83}},
  \bibinfo{pages}{2351} (\bibinfo{year}{1999}).

\bibitem[{\citenamefont{Estreicher}(2000)}]{ES00}
\bibinfo{author}{\bibfnamefont{S.~K.} \bibnamefont{Estreicher}},
  \bibinfo{journal}{Phys.\ Stat.\ sol.\ (b)} \textbf{\bibinfo{volume}{217}},
  \bibinfo{pages}{513} (\bibinfo{year}{2000}).

\bibitem[{\citenamefont{Henkelman et~al.}(2002)\citenamefont{Henkelman,
  Uberuaga, Dunham, and J\'onsson}}]{Hen02}
\bibinfo{author}{\bibfnamefont{G.}~\bibnamefont{Henkelman}},
  \bibinfo{author}{\bibfnamefont{B.}~\bibnamefont{Uberuaga}},
  \bibinfo{author}{\bibfnamefont{S.}~\bibnamefont{Dunham}}, \bibnamefont{and}
  \bibinfo{author}{\bibfnamefont{H.}~\bibnamefont{J\'onsson}},
  \bibinfo{journal}{Phys.\ Stat.\ sol.\ (b)} \textbf{\bibinfo{volume}{233}},
  \bibinfo{pages}{24} (\bibinfo{year}{2002}).

\bibitem[{\citenamefont{Munro and Wales}(1999)}]{Mu99}
\bibinfo{author}{\bibfnamefont{L.~J.} \bibnamefont{Munro}} \bibnamefont{and}
  \bibinfo{author}{\bibfnamefont{D.}~\bibnamefont{Wales}},
  \bibinfo{journal}{Phys.\ Rev.\ B} \textbf{\bibinfo{volume}{59}},
  \bibinfo{pages}{3969} (\bibinfo{year}{1999}).

\bibitem[{\citenamefont{Puska}(2000)}]{Pu00}
\bibinfo{author}{\bibfnamefont{M.~J.} \bibnamefont{Puska}},
  \bibinfo{journal}{Comp.\ Mat.\ Sc.} \textbf{\bibinfo{volume}{17}},
  \bibinfo{pages}{365} (\bibinfo{year}{2000}).

\bibitem[{\citenamefont{Mercer et~al.}(1998)\citenamefont{Mercer, Nelson,
  Wright, and Stechel}}]{Me98}
\bibinfo{author}{\bibfnamefont{J.~L.} \bibnamefont{Mercer}},
  \bibinfo{author}{\bibfnamefont{J.~S.} \bibnamefont{Nelson}},
  \bibinfo{author}{\bibfnamefont{A.~F.} \bibnamefont{Wright}},
  \bibnamefont{and} \bibinfo{author}{\bibfnamefont{E.~B.}
  \bibnamefont{Stechel}}, \bibinfo{journal}{Modelling.\ Simul.\ Mater.\ Sci.\
  Eng.} \textbf{\bibinfo{volume}{6}}, \bibinfo{pages}{1}
  (\bibinfo{year}{1998}).

\bibitem[{\citenamefont{Probert and Payne}(2003)}]{PROB03}
\bibinfo{author}{\bibfnamefont{M.~I.~J.} \bibnamefont{Probert}}
  \bibnamefont{and} \bibinfo{author}{\bibfnamefont{M.~C.} \bibnamefont{Payne}},
  \bibinfo{journal}{Phys.\ Rev.\ B} \textbf{\bibinfo{volume}{67}},
  \bibinfo{pages}{075204} (\bibinfo{year}{2003}).

\bibitem[{\citenamefont{Henkelman et~al.}(2000)\citenamefont{Henkelman,
  Uberuaga, and J\'onsson}}]{UBER00}
\bibinfo{author}{\bibfnamefont{G.}~\bibnamefont{Henkelman}},
  \bibinfo{author}{\bibfnamefont{B.~P.} \bibnamefont{Uberuaga}},
  \bibnamefont{and}
  \bibinfo{author}{\bibfnamefont{H.}~\bibnamefont{J\'onsson}},
  \bibinfo{journal}{J.\ Chem.\ Phys.} \textbf{\bibinfo{volume}{113}},
  \bibinfo{pages}{9901} (\bibinfo{year}{2000}).

\bibitem[{\citenamefont{Roberson and Estreicher}(1994)}]{ROBER94}
\bibinfo{author}{\bibfnamefont{M.~A.} \bibnamefont{Roberson}} \bibnamefont{and}
  \bibinfo{author}{\bibfnamefont{S.~K.} \bibnamefont{Estreicher}},
  \bibinfo{journal}{Phys.\ Rev.\ B} \textbf{\bibinfo{volume}{49}},
  \bibinfo{pages}{17040} (\bibinfo{year}{1994}).

\bibitem[{\citenamefont{\"O\u{g}\"ut et~al.}(1997)\citenamefont{\"O\u{g}\"ut,
  Kim, and Chelikowsky}}]{KIM97}
\bibinfo{author}{\bibfnamefont{S.}~\bibnamefont{\"O\u{g}\"ut}},
  \bibinfo{author}{\bibfnamefont{H.}~\bibnamefont{Kim}}, \bibnamefont{and}
  \bibinfo{author}{\bibfnamefont{J.}~\bibnamefont{Chelikowsky}},
  \bibinfo{journal}{Phys.\ Rev.\ B} \textbf{\bibinfo{volume}{56}},
  \bibinfo{pages}{R11 353} (\bibinfo{year}{1997}).

\bibitem[{\citenamefont{Anderson et~al.}(1996)\citenamefont{Anderson, Ham, and
  Grossmann}}]{ANDER96}
\bibinfo{author}{\bibfnamefont{F.}~\bibnamefont{Anderson}},
  \bibinfo{author}{\bibfnamefont{F.~S.} \bibnamefont{Ham}}, \bibnamefont{and}
  \bibinfo{author}{\bibfnamefont{G.}~\bibnamefont{Grossmann}},
  \bibinfo{journal}{Phys.\ Rev.\ B} \textbf{\bibinfo{volume}{53}},
  \bibinfo{pages}{7205} (\bibinfo{year}{1996}).

\bibitem[{\citenamefont{Clarck and Ackland}(1997)}]{Clarck97}
\bibinfo{author}{\bibfnamefont{S.~J.} \bibnamefont{Clarck}} \bibnamefont{and}
  \bibinfo{author}{\bibfnamefont{G.~J.} \bibnamefont{Ackland}},
  \bibinfo{journal}{Phys.\ Rev.\ B} \textbf{\bibinfo{volume}{56}},
  \bibinfo{pages}{47} (\bibinfo{year}{1997}).

\bibitem[{\citenamefont{Lento and Nieminen}(2003)}]{Len03}
\bibinfo{author}{\bibfnamefont{J.}~\bibnamefont{Lento}} \bibnamefont{and}
  \bibinfo{author}{\bibfnamefont{R.~M.} \bibnamefont{Nieminen}},
  \bibinfo{journal}{J. Phys.: Condens. Matter} \textbf{\bibinfo{volume}{15}},
  \bibinfo{pages}{4387} (\bibinfo{year}{2003}).

\bibitem[{\citenamefont{Agullo-Lopez et~al.}(1988)\citenamefont{Agullo-Lopez,
  Catlow, and Townsend}}]{Lopez88}
\bibinfo{author}{\bibfnamefont{F.}~\bibnamefont{Agullo-Lopez}},
  \bibinfo{author}{\bibfnamefont{C.}~\bibnamefont{Catlow}}, \bibnamefont{and}
  \bibinfo{author}{\bibfnamefont{P.}~\bibnamefont{Townsend}},
  \emph{\bibinfo{title}{Point Defects in Materials}}
  (\bibinfo{publisher}{Academic Press}, \bibinfo{year}{1988}).

\bibitem[{\citenamefont{Maroudas and Brown}(1993)}]{Maroud93}
\bibinfo{author}{\bibfnamefont{D.}~\bibnamefont{Maroudas}} \bibnamefont{and}
  \bibinfo{author}{\bibfnamefont{R.~A.} \bibnamefont{Brown}},
  \bibinfo{journal}{Phys.\ Rev.\ B} \textbf{\bibinfo{volume}{47}},
  \bibinfo{pages}{15562} (\bibinfo{year}{1993}).

\bibitem[{\citenamefont{Mousseau and Barkema}(1998)}]{MOU98}
\bibinfo{author}{\bibfnamefont{N.}~\bibnamefont{Mousseau}} \bibnamefont{and}
  \bibinfo{author}{\bibfnamefont{G.~T.} \bibnamefont{Barkema}},
  \bibinfo{journal}{Phys.\ Rev.\ E} \textbf{\bibinfo{volume}{57}},
  \bibinfo{pages}{2419} (\bibinfo{year}{1998}).

\bibitem[{\citenamefont{Barkema and Mousseau}(1996)}]{BAR96}
\bibinfo{author}{\bibfnamefont{G.~T.} \bibnamefont{Barkema}} \bibnamefont{and}
  \bibinfo{author}{\bibfnamefont{N.}~\bibnamefont{Mousseau}},
  \bibinfo{journal}{Phys.\ Rev.\ Lett.} \textbf{\bibinfo{volume}{77}},
  \bibinfo{pages}{4358} (\bibinfo{year}{1996}).

\bibitem[{\citenamefont{S\'anchez-Portal
  et~al.}(1997)\citenamefont{S\'anchez-Portal, Ordej\'on, Artacho, and
  Soler}}]{SAN97}
\bibinfo{author}{\bibfnamefont{D.}~\bibnamefont{S\'anchez-Portal}},
  \bibinfo{author}{\bibfnamefont{P.}~\bibnamefont{Ordej\'on}},
  \bibinfo{author}{\bibfnamefont{E.}~\bibnamefont{Artacho}}, \bibnamefont{and}
  \bibinfo{author}{\bibfnamefont{J.~M.} \bibnamefont{Soler}},
  \bibinfo{journal}{Int. \ J.\ Quant.\ Chem.} \textbf{\bibinfo{volume}{65}},
  \bibinfo{pages}{453} (\bibinfo{year}{1997}).

\bibitem[{\citenamefont{Soler et~al.}(2002)\citenamefont{Soler, Artacho, Gale,
  Garc\'ia, Junquera, and Ordej\'on}}]{Soler02}
\bibinfo{author}{\bibfnamefont{J.~M.} \bibnamefont{Soler}},
  \bibinfo{author}{\bibfnamefont{E.}~\bibnamefont{Artacho}},
  \bibinfo{author}{\bibfnamefont{J.}~\bibnamefont{Gale}},
  \bibinfo{author}{\bibfnamefont{A.}~\bibnamefont{Garc\'ia}},
  \bibinfo{author}{\bibfnamefont{J.}~\bibnamefont{Junquera}}, \bibnamefont{and}
  \bibinfo{author}{\bibfnamefont{P.}~\bibnamefont{Ordej\'on}},
  \bibinfo{journal}{J. Phys. Condens. Matter} \textbf{\bibinfo{volume}{14}},
  \bibinfo{pages}{2745} (\bibinfo{year}{2002}).

\bibitem[{\citenamefont{Malek and Mousseau}(2000)}]{Malek_ART}
\bibinfo{author}{\bibfnamefont{R.}~\bibnamefont{Malek}} \bibnamefont{and}
  \bibinfo{author}{\bibfnamefont{N.}~\bibnamefont{Mousseau}},
  \bibinfo{journal}{Phys. Rev. E} \textbf{\bibinfo{volume}{62}},
  \bibinfo{pages}{7723} (\bibinfo{year}{2000}).

\bibitem[{\citenamefont{Henkelman and J\'onsson}(1999)}]{Hen99}
\bibinfo{author}{\bibfnamefont{G.}~\bibnamefont{Henkelman}} \bibnamefont{and}
  \bibinfo{author}{\bibfnamefont{J.}~\bibnamefont{J\'onsson}},
  \bibinfo{journal}{J.\ Chem.\ Phys.} \textbf{\bibinfo{volume}{111}},
  \bibinfo{pages}{7010} (\bibinfo{year}{1999}).

\bibitem[{\citenamefont{Kresse and Hafner}(1993)}]{vasp}
\bibinfo{author}{\bibfnamefont{G.}~\bibnamefont{Kresse}} \bibnamefont{and}
  \bibinfo{author}{\bibfnamefont{J.}~\bibnamefont{Hafner}},
  \bibinfo{journal}{Phys.\ Rev.\ B} \textbf{\bibinfo{volume}{47}},
  \bibinfo{pages}{558} (\bibinfo{year}{1993}).

\bibitem[{\citenamefont{Lancz\'os}(1988)}]{Lan88}
\bibinfo{author}{\bibfnamefont{C.}~\bibnamefont{Lancz\'os}},
  \emph{\bibinfo{title}{Applied Analysis}} (\bibinfo{publisher}{Dover, New
  York}, \bibinfo{year}{1988}).

\bibitem[{\citenamefont{Troullier and Martins}(1991)}]{Trou91}
\bibinfo{author}{\bibfnamefont{N.}~\bibnamefont{Troullier}} \bibnamefont{and}
  \bibinfo{author}{\bibfnamefont{J.~L.} \bibnamefont{Martins}},
  \bibinfo{journal}{Phys.\ Rev.\ B} \textbf{\bibinfo{volume}{43}},
  \bibinfo{pages}{1993} (\bibinfo{year}{1991}).

\bibitem[{\citenamefont{Kleiman and Bylander}(1982)}]{Klei82}
\bibinfo{author}{\bibfnamefont{L.}~\bibnamefont{Kleiman}} \bibnamefont{and}
  \bibinfo{author}{\bibfnamefont{D.~M.} \bibnamefont{Bylander}},
  \bibinfo{journal}{Phys.\ Rev.\ Lett.} \textbf{\bibinfo{volume}{48}},
  \bibinfo{pages}{1425} (\bibinfo{year}{1982}).

\bibitem[{\citenamefont{Anglada et~al.}(2002)\citenamefont{Anglada, Soler,
  Junquera, and Artacho}}]{Ang02}
\bibinfo{author}{\bibfnamefont{E.}~\bibnamefont{Anglada}},
  \bibinfo{author}{\bibfnamefont{J.~M.} \bibnamefont{Soler}},
  \bibinfo{author}{\bibfnamefont{J.}~\bibnamefont{Junquera}}, \bibnamefont{and}
  \bibinfo{author}{\bibfnamefont{E.}~\bibnamefont{Artacho}},
  \bibinfo{journal}{Phys.\ Rev.\ B} \textbf{\bibinfo{volume}{66}},
  \bibinfo{pages}{205101} (\bibinfo{year}{2002}).

\bibitem[{\citenamefont{Monkhorst and Pack}(1976)}]{Mon76}
\bibinfo{author}{\bibfnamefont{H.~J.} \bibnamefont{Monkhorst}}
  \bibnamefont{and} \bibinfo{author}{\bibfnamefont{J.~D.} \bibnamefont{Pack}},
  \bibinfo{journal}{Phys.\ Rev.\ B} \textbf{\bibinfo{volume}{13}},
  \bibinfo{pages}{5188} (\bibinfo{year}{1976}).

\bibitem[{\citenamefont{Ceperley and Alder}(1980)}]{Cep80}
\bibinfo{author}{\bibfnamefont{D.~M.} \bibnamefont{Ceperley}} \bibnamefont{and}
  \bibinfo{author}{\bibfnamefont{B.~J.} \bibnamefont{Alder}},
  \bibinfo{journal}{Phys. Rev. Lett.} \textbf{\bibinfo{volume}{45}},
  \bibinfo{pages}{566} (\bibinfo{year}{1980}).

\bibitem[{\citenamefont{Perdew and Zunger}(1981)}]{Per81}
\bibinfo{author}{\bibfnamefont{J.~P.} \bibnamefont{Perdew}} \bibnamefont{and}
  \bibinfo{author}{\bibfnamefont{A.}~\bibnamefont{Zunger}},
  \bibinfo{journal}{Phys.\ Rev.\ B} \textbf{\bibinfo{volume}{23}},
  \bibinfo{pages}{5048} (\bibinfo{year}{1981}).

\bibitem[{\citenamefont{Bourgoin and Lanoo}(1981)}]{lanoo}
\bibinfo{author}{\bibfnamefont{J.}~\bibnamefont{Bourgoin}} \bibnamefont{and}
  \bibinfo{author}{\bibfnamefont{M.}~\bibnamefont{Lanoo}},
  \emph{\bibinfo{title}{Point defects is Semiconductors I and II}}
  (\bibinfo{publisher}{Springer, Berlin}, \bibinfo{year}{1981}).

\bibitem[{\citenamefont{Kelly and Car}(1992)}]{Kelly92}
\bibinfo{author}{\bibfnamefont{P.}~\bibnamefont{Kelly}} \bibnamefont{and}
  \bibinfo{author}{\bibfnamefont{R.}~\bibnamefont{Car}},
  \bibinfo{journal}{Phys.\ Rev.\ B} \textbf{\bibinfo{volume}{45}},
  \bibinfo{pages}{6543} (\bibinfo{year}{1992}).

\bibitem[{\citenamefont{Bl\"{o}chl et~al.}(1993)\citenamefont{Bl\"{o}chl,
  Smargiassi, Car, Laks, Andreoni, and Pantelides}}]{Blo93}
\bibinfo{author}{\bibfnamefont{P.}~\bibnamefont{Bl\"{o}chl}},
  \bibinfo{author}{\bibfnamefont{E.}~\bibnamefont{Smargiassi}},
  \bibinfo{author}{\bibfnamefont{R.}~\bibnamefont{Car}},
  \bibinfo{author}{\bibfnamefont{D.~B.} \bibnamefont{Laks}},
  \bibinfo{author}{\bibfnamefont{W.}~\bibnamefont{Andreoni}}, \bibnamefont{and}
  \bibinfo{author}{\bibfnamefont{S.}~\bibnamefont{Pantelides}},
  \bibinfo{journal}{Phys.\ Rev.\ Lett.} \textbf{\bibinfo{volume}{70}},
  \bibinfo{pages}{2435} (\bibinfo{year}{1993}).

\bibitem[{\citenamefont{Vineyard}(1957)}]{Vin57}
\bibinfo{author}{\bibfnamefont{H.}~\bibnamefont{Vineyard}},
  \bibinfo{journal}{J.\ Phys.\ Chem.\ Solids} \textbf{\bibinfo{volume}{3}},
  \bibinfo{pages}{121} (\bibinfo{year}{1957}).

\bibitem[{\citenamefont{Baskes et~al.}(1989)\citenamefont{Baskes, Nelson, and
  Wright}}]{Baskes}
\bibinfo{author}{\bibfnamefont{M.~I.} \bibnamefont{Baskes}},
  \bibinfo{author}{\bibfnamefont{J.~S.} \bibnamefont{Nelson}},
  \bibnamefont{and} \bibinfo{author}{\bibfnamefont{A.~F.}
  \bibnamefont{Wright}}, \bibinfo{journal}{Phys.\ Rev.\ B}
  \textbf{\bibinfo{volume}{40}}, \bibinfo{pages}{6085} (\bibinfo{year}{1989}).

\bibitem[{\citenamefont{Bracht et~al.}(1998)\citenamefont{Bracht, Haller, and
  Clark-Phelps}}]{Bracht98}
\bibinfo{author}{\bibfnamefont{H.}~\bibnamefont{Bracht}},
  \bibinfo{author}{\bibfnamefont{E.~E.} \bibnamefont{Haller}},
  \bibnamefont{and}
  \bibinfo{author}{\bibfnamefont{R.}~\bibnamefont{Clark-Phelps}},
  \bibinfo{journal}{Phys.\ Rev.\ Lett.} \textbf{\bibinfo{volume}{81}},
  \bibinfo{pages}{393} (\bibinfo{year}{1998}).

\bibitem[{\citenamefont{Tan and G\"osele}(1985)}]{Tan85}
\bibinfo{author}{\bibfnamefont{T.~Y.} \bibnamefont{Tan}} \bibnamefont{and}
  \bibinfo{author}{\bibfnamefont{U.}~\bibnamefont{G\"osele}},
  \bibinfo{journal}{Appl. Phys. A: Solids Surf.} \textbf{\bibinfo{volume}{37}},
  \bibinfo{pages}{1} (\bibinfo{year}{1985}).

\bibitem[{\citenamefont{Phillips}(2001)}]{Rob01}
\bibinfo{author}{\bibfnamefont{R.}~\bibnamefont{Phillips}},
  \emph{\bibinfo{title}{crystals, Defects and Microstructures, Modelling a
  cross Scales}} (\bibinfo{publisher}{Cambridge university press},
  \bibinfo{year}{2001}).

\bibitem[{\citenamefont{Wooten et~al.}(1985)\citenamefont{Wooten, Winer, and
  Weaire}}]{WWW85}
\bibinfo{author}{\bibfnamefont{F.}~\bibnamefont{Wooten}},
  \bibinfo{author}{\bibfnamefont{K.}~\bibnamefont{Winer}}, \bibnamefont{and}
  \bibinfo{author}{\bibfnamefont{D.}~\bibnamefont{Weaire}},
  \bibinfo{journal}{Phys.\ Rev.\ Lett.} \textbf{\bibinfo{volume}{54}},
  \bibinfo{pages}{1392} (\bibinfo{year}{1985}).

\bibitem[{\citenamefont{Cargnoni et~al.}(1998)\citenamefont{Cargnoni, Gatti,
  and Colombo}}]{CARG98}
\bibinfo{author}{\bibfnamefont{F.}~\bibnamefont{Cargnoni}},
  \bibinfo{author}{\bibfnamefont{C.}~\bibnamefont{Gatti}}, \bibnamefont{and}
  \bibinfo{author}{\bibfnamefont{L.}~\bibnamefont{Colombo}},
  \bibinfo{journal}{Phys.\ Rev.\ B} \textbf{\bibinfo{volume}{57}},
  \bibinfo{pages}{170} (\bibinfo{year}{1998}).

\end{thebibliography}
\end{document}